\documentstyle[12pt]{article}
\newcommand{\comma}{\;\;\;\; ,}
\newcommand{\period}{\;\;\;\; .}

\newcommand{\eq}{\; = \;}
\newcommand{\sep}{\; , \;\;\;}
\newcommand{\be}{\begin{equation}}
\newcommand{\bd}{\begin{displaymath}}
\newcommand{\ee}{\end{equation}}
\newcommand{\ed}{\end{displaymath}}
\newcommand{\ba}{\begin{eqnarray}}
\newcommand{\ea}{\end{eqnarray}}

\newcommand{\Trace}{{\rm Trace} \,}

\title{Planar lattice gases with nearest-neighbour exclusion.}

\author{ R.J. Baxter \\
{\protect \small Theoretical Physics, I.A.S. and School of Mathematical
Sciences}\\
{\protect \small  The Australian National University,
 Canberra, A.C.T. 0200, Australia  }
}
\date{}

\begin{document}

\maketitle

\abstract{We discuss the hard-hexagon and hard-square 
problems, as well as the corresponding problem on the 
honeycomb lattice. The case when the activity is unity
is of interest to combinatorialists, being the problem of counting 
binary matrices with no two adjacent 1's. For this case
we use the powerful corner transfer matrix method to 
numerically evaluate the partition function per site, density and
some near-neighbour correlations to high accuracy. In particular for the square lattice 
we obtain the partition function per site to 43 decimal places.
}

\section{Introduction}
\subsection*{Partition function of hard squares}
In recent years there has been interest amongst combinatorialists
in the problem of counting the number of ``legal matrices'' \cite{Calkin98,FinchWeb,Finch98,
Engel90}. These are matrices whose elements are either 0 or 1. Two elements are said to 
be `adjacent' 
if they lie in positions $(i,j)$ and $(i+1,j)$, or if they lie in positions $(i,j)$
 and $(i,j+1)$.
No two 1's are allowed to be adjacent.

In statistical mechanics this is known as the \lq hard squares\rq problem: 
how many  ways are there of putting particles on a square lattice of $N$ sites
so that no two share the same site, or are adjacent. (An economical way of 
formulating this rule is to say that there is at most one particle on every 
edge.)

In statistical mechanics we usually generalize this problem by specifying 
the number of particles. Thus we might define $g(m,N)$ to be the number of ways of
putting m particles on a given lattice of N sites, subject to the above 
exclusion rule. This is the \lq canonical' partition function. The 
\lq grand canonical partition function' is
\be \Xi_N(z) \eq \sum_m  g(m,N) z^m \comma \ee
$z$ being an arbitrary real (or complex) variable. For $z$ real and positive
we expect on physical grounds that the limit 
\be \kappa (z ) \eq  \lim_{N \rightarrow \infty } \left[ \Xi_N(z) \right] ^{1/N}
\ee
will exist and be independent of the shape of the lattice, so long as
the shape is not pathological - it must certainly  become infinite in all 
directions.

It follows that what combinatorialists are interested in is $\Xi_N(1)$
and $\kappa (1)$. While the function $\kappa (z)$ has been extensively 
investigated, by series expansions and numerically \cite{BaxterEnting80,GauntFisher65, 
PearceSeaton88,RunnelsCoombs66}, less attention 
has been paid by statistical mechanics to its value at $z = 1$. Estimates for $\kappa (1)$
 of 1.5030 and 1.503048082 are given in \cite{MetcalfYang78} and \cite{Milosevic89}, and 
lower and upper bounds of 1.503263 and 1.517513 in \cite{Markley81}.

Amongst combinatorialists, Engel \cite{Engel90} obtained the estimate
1.50304808, and Calkin and Wilf \cite{Calkin98} have recently obtained 
1.5030480824753323, with rigorous lower and upper bounds 1.503047782 and
1.5035148. This estimate has been refined by McKay to 1.50304808247533226 
\cite{McKay}.

Here we use the corner transfer matrix method to calculate  $\kappa (1)$ and 
various related quantities and obtain the result
~~~1.5030480824753322643220663294755536893857810. 
Based on the observed convergence of the sequence of successive approximations 
(and on the expected error of ${\cal A}_{49}^4$),
we believe this result to be correct to the 43 decimal places given.

\subsection*{Related models and quantities}

For the corresponding problem on the triangular lattice, 
known as \lq hard hexagons\rq , Metcalf and Yang 
in 1978 \cite{MetcalfYang78} conjectured that $ \log \kappa (1)  = 1/3$.
This intriguing conjecture prompted Baxter and Tsang \cite{BaxterTsang80}
to obtain more accurate numerical values, using the corner transfer matrix 
method \cite{BaxterCTM81}. They disproved Metcalf and Yang's conjecture, but
as a result the author stumbled on the fact that the problem could be solved 
exactly by the \lq Yang-Baxter'
method (C.N. Yang, rather than his brother C.P.). \cite{Baxter80, 
BaxterCitnClassic90}

For the square and honeycomb lattices, the problem has not been solved
exactly and the suggestive clues of hard hexagons are missing. Even so,
the method provides a powerful way 
of calculating $\kappa (1)$  to considerable numerical precision.  We do 
this here, obtaining it to 42 and 39 decimal places for the  square and 
honeycomb lattices, respectively. For completeness, we also evaluate $\kappa (1)$ 
for the triangular lattice, using the known exact result, and give it here to 55 decimal places.

A related quantity is the density, or probability that a given site is
occupied:
\be \label{rhokappa}
\rho (z) \eq \Xi_N(z)^{-1} \sum_m   (m/N) g(m,N) z^m \eq 
z \frac{d}{dz} \ln \kappa (z) \comma \ee
also expected to tend to a limit for a large lattice.
This is a \lq single-site' property and is also readily calculated
by the corner transfer matrix method, so we present numerical values
for $\rho (1)$ here, along with some next-nearest neighbour correlation
results.

It should be noted that the models are expected to undergo a phase transition
from a disordered fluid phase to an ordered solid crystalline phase at
$z = 11.09017..., 3.7962...$, $7.92...$, $\rho = 0.27639..., 0.368..., 0.422...$,
for the triangular, square and honeycomb lattices, respectively. 
\cite{Baxter80,BaxterEnting80, RunnelsCoombs67}. For each lattice, these critical values of $z$
are considerably greater than $z = 1$, so the gas is well and truly in the fluid
phase and we expect successive finite truncations of the corner transfer matrix 
equations (with no sub-lattice symmetry breaking) to converge rapidly, and indeed
we observe this. The method converges more slowly as $z$  
approaches the critical value.

\section{Low density series expansions}

The  series expansions were obtained for these models in 
the sixties \cite{Gaunt67,GauntFisher65,Sykes65} to orders $z^8, z^{13}, z^{13}$, 
respectively. They have since been extended, e.g. in \cite[p. 408]{BaxBook82}, 
\cite{BaxterEnting80}. Here we give each to order $z^{13}$:
{\samepage
\bd {\rm triangular:}\; \;  \kappa \eq 1 + z - 3\,{z^2} + 16\,{z^3} - 
106\,{z^4} +  789\,{z^5} - 6318\,{z^6} + \ed
\bd  53198\,{z^7} -  464673\,{z^8} + 4174088\,{z^9} - 
38332500\,{z^{10}} + 358371345\,{z^{11}} - \ed
\bd  3400238419\,{z^{12}} + 
   32664062740\,{z^{13}} + 
  {\rm O}(z^{14}) \ed
}

{\samepage
\bd   {\rm square:} \; \;    
\kappa \eq 1 + z - 2\,{z^2} + 8\,{z^3} - 40\,{z^4} + 
   225\,{z^5} - 1362\,{z^6} + \; \; \; \; \; \; \; \ed
\be \label{series} 8670\,{z^7} -  57253\,{z^8} + 388802\,{z^9} - 2699202\,{z^{10}} +   
19076006\,{z^{11}} - \ee
\bd 136815282\,{z^{12}} + 993465248\,{z^{13}} + 
  {\rm O}(z^{14}) \ed
}

{\samepage
\bd {\rm honeycomb:} \; \; \kappa^2 \eq  1 + 2\,z - 2\,{z^2} + 6\,{z^3} - 23\,{z^4} + 100\,{z^5} - 
   469\,{z^6} +\ed
\bd  2314\,{z^7} - 11841\,{z^8} + 62286\,{z^9} - 
   334804\,{z^{10}} + 1831358\,{z^{11}} - \ed \bd
  10162679\,{z^{12}} + 57080840\,{z^{13}} + 
  {\rm O}(z^{14}) \ed
}

\section{Corner transfer matrix equations}

Label the sites of the lattice $i = 1, \ldots , N$. With site $i$ associate
an occupation number (or spin) $\sigma_i$, with value 0 if the site is empty,  
1 if it is full.  Then for all three lattices the partition function is
\be
\Xi_N (z) \eq \sum \prod_{i} z^{\sigma_i} \, \prod_{<ij>} e(\sigma_i, \sigma_j) \comma \ee
where the sum is over all values 0 or 1 of all the occupation numbers 
(without restriction), the second product is over all edges $<i,j>$ of the lattice,
and \be e(0,0) = e(0,1) = e(1,0) = 1  \sep e(1,1) = 0 \period \ee

For the {\it triangular} lattice the corner transfer matrix equations are given in
\cite{BaxterTsang80}. They are
\ba \label{ctm1}
\sum_b F(a,b) A^2(b) F(b,a) & = & \xi \, A^4(a) \comma \nonumber \\
\sum_c w(a, b, c) F(a,c) A(c) F(c,b) & = & \eta \, A(a) F(a,b) A(b) \comma \\ 
\kappa  & = & \eta^2 /\xi \period \nonumber \ea
Here $a, b, c$ are site occupation numbers, taking the values 0 or 1, and
$w(a, b, c)$ is the weight function of a triangular face of the lattice, which for our 
hard molecule model is
\bd w(a, b, c)  \eq  z^{(a+b+c)/6} e(a, b) e(b, c) e(c, a) \period \ed

For the {\it square} lattice the corner transfer matrix equations are given in
\cite{VarApprox78} and \cite{BaxterTsang80}. Specializing to the translation-invariant case 
(with no symmetry breaking), they are
\ba \label{ctm2}
\sum_b F(a,b) A^2(b) F(b,a) & \eq  & \xi \, A^2(a) \comma \nonumber \\
\sum_{c,d} w(a, c, d, b) F(a,c) A(c) F(c,d) A(d) F(d, b) & \eq & \eta A(a) F(a,b) A(b) \comma \\
\kappa & \eq & \eta / \xi \comma \nonumber \ea
the face weight function being
\bd w(a, b, c, d) \eq z^{(a + b + c + d)/4} e(a, b) e(b, c) e(c, d) e(d,a) 
\period \ed

For the {\it honeycomb} lattice, we found it convenient to transform the model to one on the 
triangular lattice by dividing it into two sub-lattices and summing over all 
the spins on one sub-lattice. (This is a decimation transformation.) 
The resulting triangular lattice model has different face weights $w_1, w_2$ for 
up-pointing and 
down-pointing triangles, respectively:
\bd w_1(a, b, c)  \eq  z^{(a+b+c)/3} + z \delta_{a0} \delta_{b0} \delta_{c0} 
\sep
 w_2(a, b, c )  \eq  1  \comma \ed
and the equations (\ref{ctm1}) generalize to
\ba \label{ctm3}
\sum_b F_i(a,b) A_i(b) A_j(b) F_j(b,a) & = & \xi \, A_j(a) A_i(a) A_j(a) A_i(a) \comma \nonumber \\
\sum_c w_i(a, b, c) F_i(a,c) A_i(c) F_i(c,b) & = & \eta_i \, A_j(a) F_j(a,b) A_j(b) \comma \\ 
\kappa  & = & (\eta_1 \eta_2 /\xi)^{1/2} \comma \nonumber \ea
for (i,j) = (1,2) and (2,1). Here $\kappa$ is the partition function per site of the 
original honeycomb lattice.

In each case $A(a)$ is the corner transfer matrix centred on a site of occupation 
number $a$. Define the matrix
\be
M(a) =  A^6(a) \sep  A^4(a) \sep  [A_1(a) A_2(a)]^3 \comma \ee
for the three lattices, respectively. Then the density is
\be \label{dens}
 \rho \eq \Trace M(1)/[\Trace M(0) + \Trace M(1)] \period \ee
Similarly, $F(a,b)$ is the `half-row' transfer matrix associated with an edge 
with occupation numbers $a, b$. For this reason $F(1,1)$ is the zero matrix for 
the first two lattices. (For the honeycomb lattice the decimation transformation 
means that $F_i(a,b)$ is associated with next-nearest neghbour sites $a, b$, so 
$F_i(1,1)$ is not zero.) Other local occupation probabilities can be obtained by building up
a face or faces of the lattice. For instance, for the square lattice the
probability that two next-nearest-neighbour (i.e. diagonally adjacent) sites are
simultaneously occupied is
\be \rho_{nn} \eq   \frac{w(1, 0, 1, 0) \Trace [A(1) F(1,0) A(0) F(0,1) A(1) F(1,0)
 A(0) F(0,1)]}{
  \sum_{a,b,c,d} w(a, b, c, d) \Trace [A(a) F(a,b) A(b) F(b,c) A(c) F(c,d)
 A(d) F(d,a)]} \period \ee

It can be convenient to define the larger matrices
\be \label{largematrices}
{\cal A} \eq \left( \begin{array}{cc} A(0) & 0 \\0 & A(1) \end{array} \right)
\sep
{\cal F} \eq \left( \begin{array}{cc} F(0,0) & F(0,1) \\F(1,0) & F(1,1) \end{array} 
\right) \ee
\bd
{\cal S} \eq \left( \begin{array}{cc} 0 & 0 \\0 & 1 \end{array} \right) \sep
{\cal M} \eq  \left( \begin{array}{cc} M(0) & 0 \\0 & M(1) \end{array} \right) \period \ed
(Generalizing as needed for the honeycomb lattice.) 

These extended  matrices $\cal A$, $\cal F$, $\cal M$ are real and symmetric. The corner transfer matrix equations above define
these matrices to within irrelevant similarity transformations: it is convenient to remove this 
arbitrariness by choosing $\cal A$ and $\cal M$ to be diagonal.

The equations are exact in the limit when the matrices are infinite-dimensional.
Finite-dimensional truncations provide a sequence of rapidly converging approximations
and the techniques needed to handle these are explained in 
\cite{VarApprox78,BaxterCTM81, BaxterTsang80}. For each lattice a column $k, b$ of
the second equation plays the role of an eigenvalue equation, the  eigenvalue being the diagonal 
element $k, b$ of $\cal A$, and the eigenvector being column $k, b$ of $\cal F$. The first 
equation fixes the normalization of the eigenvector. In any finite truncation there are more 
eigenvalues than needed: one selects those giving the larger eigenvalues of $\cal M$.
One can normalise the diagonal matrix $\cal M$ to have maximum element unity, the diagonal
elements being arranged in numerically decreasing order. These elements do not change dramatically
as one increases dimensionality, so one can regard the truncation where $\cal M$ 
has dimensionality $r$ as providing an approximation for the largest $r$ elements
of the true infinite-dimensional matrix $\cal M$. The magnitude of the largest 
element omitted provides a measure (in practice it seems a somewhat over-confident 
measure) of the accuracy of any given truncation.

For each lattice, and for any finite truncation, the CTM equations can be obtained from a variational principle 
for $\kappa$.

\section{Simple approximations}
\subsection*{Bethe approximation}

A simple approximation for lattice models is to replace the lattice by a 
Bethe lattice (i.e. a Cayley tree with surface effects excluded) of the same 
coordination number $q$. Specializing the formula given in Chapter 4 of 
\cite{BaxBook82},
we get
\be \label{Bethe1}
\kappa \eq (1+m)/(1-m^2)^{q/2} \sep \rho \eq m/(1+m) \comma 
\ee
where $m$ is related to $z$ by
\be \label{Bethe2} z \eq m/(1-m)^q \period \ee
This approximation is self-consistent in that the relation (\ref{rhokappa})
is satisfied. When $z$ is small, so is $m$ and we readily 
obtain the series expansions for the three lattices (with $q = 6, 4, 3$, 
respectively):
\ba \label{Betheapprox}
\kappa & \eq & 1 + z - 3\,{z^2} + 18\,{z^3}  + {\rm O}(z^4) \comma \nonumber \\
\kappa & \eq & 1 + z - 2\,{z^2} + 8\,{z^3} - 41\,{z^4} + 
  {\rm O}(z^5) \comma \\
\kappa^2 & \eq & 1 + 2\,z - 2\,{z^2} + 6\,{z^3} - 23\,{z^4} + 
    100\,{z^5} - 470\,{z^6} + {\rm O}(z^7) \period \nonumber \ea

Let $q'$ be the number of edges round a face of the lattice (this is 
the coordination number of the dual lattice), so $q' = 3, 4, 6$ for the three 
lattices, respectively. Comparing (\ref{Betheapprox}) with (\ref{series}), 
we see that for each lattice the Bethe approximation fails at order $z^{q'}$,
which is the order at which a circuit first makes a contribution to the 
partition function.

\subsection*{Kramers-Wannier approximation}

A more accurate approximation, which at least in some circumstances
is related to that of Kramers and Wannier \cite{KramersWannier41}, is to take
the corner transfer matrices $A(a), F(a,b)$ in  (\ref{ctm1}) - (\ref{ctm3}) to be 
one-by-one matrices. Then ${\cal A}, {\cal F}$ are two-by-two matrices. 
For the {\it triangular} lattice this gives
\be \label{KWtri}
\kappa \eq (1-m)^3/(1-2 m)^2 \sep \rho \eq m/(1+m) \comma \ee
where $\cal M$ has the diagonal entries $1, m$ and
\bd z = m(1-m)^6/(1-2 m)^6 \period \ed
Expanding, this gives
\be \label{KW1}
\kappa \eq 1 + z - 3\,{z^2} + 16\,{z^3} - 106\,{z^4} + 
    789\,{z^5} - 6319\,{z^6} + {\rm O}(z^{7}) \period \ee

For the {\it square} lattice,
\be \label{KWsqr}
\kappa \eq (1+t)^2/(1+t-t^2) \sep \rho \eq m/(1+m) \comma \ee
where now $t, m, z$ are related by
\bd
z = t(1+t)^4/(1+t-t^2) \sep m = t/(1+t-t^2) \comma \ed
and $t$ is small when $z$ is. This yields the expansion

{\samepage
\be \label{KW2}
\kappa = 1 + z - 2\,{z^2} + 8\,{z^3} - 40\,{z^4} + 
    225\,{z^5} - 1362\,{z^6} + 8670\,{z^7} - \ee
\bd 57254\,{z^8} + 
   388830\,{z^9} +  {\rm O}(z^{10}) \period \ed
}

For the {\it honeycomb} lattice, truncating $A_i (a)$ and $F_i (a,b)$ in (\ref{ctm3})
to one-by-one, we obtain eight distinct equations for $\xi, \eta_1, \eta_2$ and 
the elements of the matrices. We have not succeeded in significantly simplifying 
the solution of these eqns, but we do note that to leading order in
$z$ we can naturally choose,
\bd {\cal A}_1 \eq \left(\begin{array}{cc} 1 & 0 \\ 0 & 1 \end{array} \right)
\sep {\cal A}_2 \eq \left(\begin{array}{cc} 1 & 0 \\ 0 & z^{1/3} \end{array} 
\right) \ed
and
\bd {\cal F}_1 \eq {\cal F}_2 \eq \left(\begin{array}{cc} 1 & z^{1/3} \\ 
z^{1/3} & z^{2/3} \end{array} \right) \period \ed
Without loss of generality we can fix ${\cal A}_1$ to be the unit matrix, and 
normalise ${\cal A}_2, {\cal F}_1, {\cal F}_2$ to have top-left element 
equal to unity. We can then iteratively solve the eight equations for the eight
remaining unknowns to successively higher orders, obtaining:
 
{\samepage
\bd \kappa^2 \eq 1 + 2\,z - 2\,{z^2} + 6\,{z^3} - 23\,{z^4} + 100\,{z^5} - 
  469\,{z^6} + \ed
\be \label{KW3} 2314\,{z^7} - 11841\,{z^8} + 62286\,{z^9} - 
  334804\,{z^{10}} + 1831358\,{z^{11}} - \ee
\bd
  10162680\,{z^{12}} + 57080872\,{z^{13}} + 
  {\rm O}(z^{14}) \ed
}

We note that for each lattice this `Kramers-Wannier' approximation is more accurate
than the Bethe approximation, being correct for orders up to (but not including)
$z^{2 q'}$.

One obtains successively more accurate approximations by taking the matrices
$A, F$ to be larger. This fact is used in \cite{BaxterEnting79} and \cite{BaxterEnting80}
to obtain  series expansions. Here we use it to obtain accurate numerical values
of $\kappa$ and $\rho$  for $z = 1$. If $A(0)$ is of size $n_0$ by $n_0$ and $A(1)$ 
is $n_1$ by $n_1$, then $F(0,1)$ is $n_0$ by $n_1$ and $\cal A$, $\cal F$ are 
$n$ by $n$, where
\be n \eq n_0 + n_1 \period \ee
From now on we re-arrange the combined matrices $\cal A$, $\cal F$,
$\cal S$, $\cal M$ so that the  entries of the diagonal matrix $\cal A$ (${\cal A}_1
{\cal A}_2$ for the honeycomb lattice) are in numerically decreasing order. Then $\cal S$
is a diagonal matrix with entries $0$ or $1$, corresponding to whether the centre site 
is empty or full for that state. The formula (\ref{dens}) can still be written
\be \rho \eq {\rm Trace}\; {\cal S} {\cal M}/{\rm Trace}\;{\cal M} \period \ee

\section{Values of $\kappa$ and $\rho$ for $z = 1$}

  Throughout this section we take $z$ to be 1. For all three lattices 
the `Kramers-Wannier' approximation has $n_0, n_1, n = 1, 1, 2$. The next 
approximation has $n_0, n_1, n = 2, 1, 3$. We present the numerical results 
obtained for the Bethe approximation and these two simple corner matrix 
approximations, together with those for $n_0, n_1, n = 3, 2, 5$. It is apparent 
that the results are converging rapidly towards a limit. For the square lattice we
have continued the sequence of approximations to $n = 48$, i.e. we have kept the largest 48
diagonal elements of $\cal A$. Then $n_0, n_1 = 29, 19$. (In fact, since 
writing this paper we 
have extended the square lattice calculation to $n = 60$, working to 55-digit accuracy, 
which has increased our confidence in the numerical results for this case.)  Similarly for the honeycomb lattice
we have continued to $n = 20$, keeping the largest 20 elements of 
${\cal A}_1 {\cal A}_2$, with $n_0, n_1 = 11, 9$.

For the benefit of anyone who wants to repeat these calculations, in the Appendix we have
given the diagonal elements of $\cal A$ and ${\cal A}_1 {\cal A}_2$ for the square and honeycomb lattices with
$n = 48$ and $n = 20$, respectively. In each case we have included the next-largest eigenvalue,
which is  obtained during the iterative solution procedure, but not actually used in the 
equations (\ref{ctm2}), (\ref{ctm3}). This provides a measure of the accuracy of the approximation.
The values of these elements change as one changes $n$, but not by more than one per cent.
For instance, the fifth eigenvalue of $\cal A$ for the square lattice occurs first when 
$n = 5$. For this and the $n = 6, 7, 8, 48 $ truncations it is
0.0077696234, 0.0077684372, 0.0077696114, 0.0077696231, 0.0077696235. Any given element of $\cal A$ or 
$\cal F$ tends quite quickly to its limiting $n = \infty$ value as $n$ increases, once $n$ is big enough for 
the element to occur at all in the truncation. 

For the triangular lattice the results for $n = 2, 3, 5$ are given in 
\cite{BaxterTsang80}. We did not pursue the corner transfer matrix
numerical calculations any further for this lattice as  (a) the exact result 
is given in \cite{Baxter80}, in chapter 14 of \cite{BaxBook82} and (in algebraic form) 
in \cite{Joyce88},   (b) the fact 
that it is exactly solvable is connected with the fact that $\cal A$
has degenerate eigenvalues, at any rate in the infinite-matrix-size limit.
These degeneracies actually complicate the numerical calculation, producing extra degrees of freedom 
that need to be fixed. Unfortunately we observed no such degeneracies for 
the square and honeycomb lattices, so we have no reason to suppose that 
they are solvable 
by the means used for the triangular case.

The `$n = \infty$' results we quote for the triangular lattice are obtained 
from the formulae (14.1.20) and (14.5.14) of \cite{BaxBook82}, using the 
solution of (14.1.18) to 55 significant digits:
\be x = - 0.2549635631051309947933138248965459184034888000019327401 \period \ee

The solutions of (\ref{Bethe1}) for the triangular, square and honeycomb lattices
(with $q = 6, 4, 3$, respectively) are $m = 0.2219104, 0.2755080, 0.3176722$.
  
The solution of (\ref{KWtri})
is $m = 0.1932944673$, that of (\ref{KWsqr}) is $t = 0.3598273461$, while for 
the honeycomb lattice the `Kramers-Wannier' two-by-two solution of (\ref{ctm3}), 
(\ref{largematrices}) is $\xi = 1.2736933548$, 
$\eta_1 = 2.4537675065$,  $\eta_2 = 1.2413585145$,

\vspace{0.2cm}
\bd
{\cal S} \eq \left( \begin{array}{cc} 0 & 0 \\0 & 1 \end{array} \right) \sep
{\cal A}_1 \eq \left( \begin{array}{cc} 1 & 0 \\0 & 1 \end{array} \right) \sep
{\cal A}_2 \eq \left( \begin{array}{cc} 1 & 0 \\0 & 0.6839628103 \end{array} 
\right) \comma \ed
\bd {\cal F}_1 \eq \left( \begin{array}{cc} 1 & 0.6736226737 \\ \! \! 0.6736226737 &
 0.4800087072 \end{array} \right)  , \;
{\cal F}_2 \eq \left( \begin{array}{cc} 1 & 0.5940391501 \\ \! \! 0.5940391501 &
 0.5960317063 \end{array} \right) \; . \ed
\vspace{0.2cm}

The values of $\kappa$ and $\rho$ obtained  are given 
in Tables 1 and 2, respectively. They are given to sufficient accuracy to see the 
convergence of the successive approximations.The deviation of an approximation 
from the extrapolated limit of the sequence (or what is almost the same: 
the deviation from the next approximation) is indeed found to be of the order of
 (or at most two or three orders greater than) the 
magnitude of the largest diagonal element of $\cal M$ omitted. 
This gives us some confidence that the sequence is indeed converging to the correct
value, and the accuracy of the best approximation.

\begin{table}
\begin{tabular}{|c|c|lll|}
\hline
approx. & n & ~~~~triangular & ~~~~square & ~~~~honeycomb \\
\hline
{\rm Bethe} &  &   1.42178 &                   1.49365   &        1.545634\\
KW &  2 &   1.395217  &                 1.502928407 &      1.546439299 \\
 & 3 &   1.3954858818              & 1.5030479990 &     1.546440707581\\
 & 5 &      1.3954859724543 &     1.50304808247401 &    1.54644070878756097\\
 & $ \infty$ &1.3954859724791398 & 1.5030480824753323 &   1.546440708787561419 \\
\hline
\end{tabular} 
\caption{{\protect{\footnotesize Values of $\kappa$ for the three planar lattices in the Bethe and 
 Kramers-Wannier (KW) approximations,  and using higher $n$ by $n$  
corner transfer matrix truncations. The extrapolated values (exact to the 
accuracy given) are in the last row.}}}
\end{table}

In fact we have obtained $\kappa$ and $\rho$ to considerably more accuracy than we are able to 
display in Tables 1 and 2. For the triangular, square and honeycomb lattices they are:

{\samepage
\ba 
\kappa & \eq &
1.3954859724793027352295006635668880689541037281446611908 \; \; \; \; (55) \nonumber\\
 & \eq & 1.5030480824753322643220663294755536893857810  \; \; \; \;  (43)  \nonumber\\
 & \eq & 1.54644070878756141848902270530472278026 \; \; \; \; (38) \nonumber\ea
}

{\samepage
\ba \rho & \eq &
 0.1624329213974881529255929066818976201010622684352448332 \; \; \; \; (55)  
\nonumber \\  
& \eq & 0.22657081546271468894199226347129902640080 \; \; \; \;  (41)  \nonumber \\
& \eq &   0.2424079763616482188205896378263422541  \; \; \; \;  (37) 
\nonumber \ea
}

The bracketted numbers are the number of decimal places given: for the triangular and square
lattices we believe the results are accurate to this number of digits. For the honeycomb lattice the last two 
or three digits should be treated with caution.

\begin{table}
\begin{tabular}{|c|c|lll|}
\hline
approx. & n & ~~~~triangular & ~~~~square & ~~~~honeycomb \\
\hline
{\rm Bethe} & &      0.18161 &         0.215999 &            0.241086  \\
KW &  2 &      0.161984 &        0.226281 &            0.242402036\\
& 3 &      0.162432600  &    0.2265703831 &        0.242407970823\\
& 5 &      0.162432921264  & 0.226570815452312 &   0.24240797636164382\\
& $\infty$ & 0.162432921397  & 0.22657081546271469 & 0.2424079763616482188 \\
\hline
\end{tabular} 
\caption{{\protect{\footnotesize The values of $\rho$ for the three lattices 
in the various approximations.}}}
\end{table}

\subsection*{Next-nearest neighbour correlation}

For all these models there is if course zero probability that two adjacent
sites are simultaneously occupied, i.e $\langle \sigma_i \sigma_j \rangle = 0$ if sites $i, j$
are adjacent. For the square and honeycomb lattices, our corner transfer matrix 
approximations also immediately give the probability that two next-nearest-neighbour sites 
are simultaneously occupied. Yet higher correlations can be obtained by constructing 
the partition function of a small lattice, using the $A$ and $F$ matrices 
appropriately to include the contributions of the boundary sites and edges. For the 
{\it square} lattice, with $i, j, k, m$ being 
the sites shown in Fig.1,
 we find (to 38 decimal places)

\begin{figure}
\begin{picture}(350,150)(-50,20)
\put(75,50) {\line(1,0){200}}
\put(75,100) {\line(1,0){200}}
\put(75,150) {\line(1,0){200}}
\put(100,25) {\line(0,1){150}}
\put(150,25) {\line(0,1){150}}
\put(200,25) {\line(0,1){150}}
\put(250,25) {\line(0,1){150}}
\put(205,39) {i}
\multiput(100,50)(50,0){4} {\circle*{5}}
\multiput(100,100)(50,0){4} {\circle*{5}}
\multiput(100,150)(50,0){4} {\circle*{5}}
\put(255,89) {j}
\put(205,89) {p}
\put(155,89) {m}
\put(155,89) {m}
\put(205,139) {k}
\end{picture}
\caption{{\protect{\footnotesize Four sites $i, j, k, m$ of the square lattice. }}}
\end{figure}
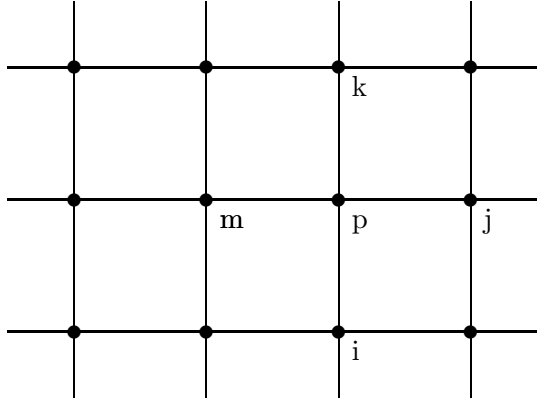

{\samepage
\ba
\rho_2 \eq \langle \sigma_i \sigma_j \rangle & \eq & 0.08198200258221599221104301382489246711 
   \nonumber \\
{\rho_2}' \eq  \langle \sigma_i \sigma_k \rangle & \eq &   0.06967964946634484029326972372242595563
\nonumber \\
\rho_3 \eq  \langle \sigma_i \sigma_j \sigma_k \rangle & \eq &  0.03024840234446871928473381972162222576 
\nonumber \\
\rho_4 \eq  \langle \sigma_i \sigma_j \sigma_k \sigma_m \rangle & \eq & 
 0.01313119289260936136017735696986128175
\period \nonumber \ea    }
This information is sufficient to calculate all correlations between the four sites 
$i, j, k, m$. Since the centre site $p$ is connected to the rest of the lattice only 
via $i, j, k, m$, we can use a ``star-square'' relation to calculate
$\langle {\sigma}_p \rangle$ from these correlations 
\cite[eqn. 80]{Fisher59}, \cite{Enting77, Fisher63a, Fisher63b}. The result is the 
relation
\be 
6 \rho - 4 \rho_2 - 2 {\rho_2}' + 4 \rho_3 - \rho_4 \eq 1 \period \ee
This is a useful check on our numerics: it is indeed satisfied to the available
accuracy.

For the {\it honeycomb} lattice, with $i, j, k, m$ being the sites shown in Fig. 2,
to 33 decimal places we obtain

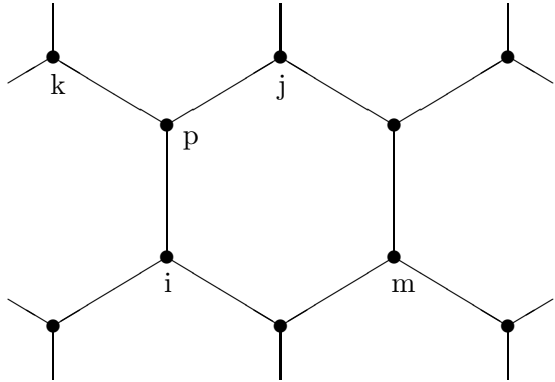
\begin{figure}
\begin{picture}(350,150)( -15,-10)
\put(250,25) {\line(0,1){50}}
\put(250,75) {\circle*{5}}
\put(250,25) {\circle*{5}}
\put(293,101) {\circle*{5}}
\put(207,101) {\circle*{5}}
\put(293,-1) {\circle*{5}}
\put(207,-1) {\circle*{5}}
\put(250,75) {\line(5,3){43}}
\put(250,75) {\line(-5,3){43}}
\put(250,25) {\line(5,-3){43}}
\put(250,25) {\line(-5,-3){43}}
\put(164,25) {\line(0,1){50}}
\put(164,75) {\circle*{5}}
\put(164,25) {\circle*{5}}
\put(121,101) {\circle*{5}}
\put(121,-1) {\circle*{5}}
\put(164,75) {\line(5,3){43}}
\put(164,75) {\line(-5,3){43}}
\put(164,25) {\line(5,-3){43}}
\put(164,25) {\line(-5,-3){43}}
\put(293,101) {\line(0,1){20}}
\put(207,101) {\line(0,1){20}}
\put(121,101) {\line(0,1){20}}
\put(293,-21) {\line(0,1){20}}
\put(207,-21) {\line(0,1){20}}
\put(121,-21) {\line(0,1){20}}
\put(293,-1) {\line(5,3){17}}
\put(121,-1) {\line(-5,3){17}}
\put(293,101) {\line(5,-3){17}}
\put(121,101) {\line(-5,-3){17}}
\put(163,12) {i}
\put(170,68) {p}
\put(206,87) {j}
\put(120,87) {k}
\put(249,12) {m}
\end{picture}
\caption{{\protect{\footnotesize Four sites $i, j, k, m$ of the honeycomb lattice. }}}
\end{figure}

\ba \rho_2 \eq \langle  \sigma_i \sigma_j \rangle & \eq & 0.079618322060174417313883811021777   
 \nonumber \\
\rho_3 \eq \langle \sigma_i \sigma_j \sigma_k \rangle & \eq &   0.026815084372282157838703243933620   
 \nonumber \\
{\rho_3}' \eq \langle \sigma_i \sigma_j \sigma_m \rangle & \eq &  0.032467404849412503526876270338554  
 \period \ea
The site $p$ is connected to the rest of the lattice only via 
$i, j, k$, so this time we can use the
 ``star-triangle'' relation to obtain
\be 5 \rho - 3 \rho_2 + \rho_3 \eq 1 \comma \ee
which is indeed satisfied to the available accuracy.

\section{Acknowledgements}
The author thanks Steven Finch and Brendan McKay for correspondence with regard 
to this problem, and for informing him of relevent work.

{\renewcommand{\arraystretch}{0.7}
\begin{table}
\begin{tabular}{|rclc|rcl|}
\hline
$i$ & $s_i$ & ~~$a_i$ & & $i~~$ & $s_i$ & ~~~~$a_i$ \\
\hline
~~~~~1    & 0 & 1.0 & &
~~~~~ 25   & 0 & $0.284464  	\times 10^{-7}$ \\
2    & 1 & $0.735716$ & &
26   & 1 & $0.219536  	\times 10^{-7}$ \\ 
3    & 0 & $0.107549$ & & 
27   & 0 & $0.954738  	\times 10^{-8}$ \\ 
4    & 1 & $0.165637 \times 10^{-1}$~~~ & & 
28   & 0 & $0.671511  	\times 10^{-8}$ \\	 
5    & 0 & $ 0.776962 \times 10^{-2}$ & & 
29   & 0 & $0.570949  	\times 10^{-8}$ \\
6    & 1 & $ 0.100064 \times 10^{-2}$ & & 
30   & 1 & $0.480772  	\times 10^{-8}$ \\ 
7    & 0 & $ 0.737188  	\times 10^{-3}$ & & 
31   & 0 & $0.307834  	\times 10^{-8}$ \\ 
8    & 0 & $ 0.260396  	\times 10^{-3}$ & & 
32   & 1 & $0.258802  	\times 10^{-8}$ \\ 
9    & 1 & $ 0.772256  	\times 10^{-4}$ & & 
33   & 1 & $0.157367  	\times 10^{-8}$ \\
10   & 0 & $ 0.593047  	\times 10^{-4}$ & & 
34   & 0 & $0.150602  	\times 10^{-8}$ \\
11   & 0 & $ 0.133328  	\times 10^{-4}$ & & 
35   & 0 & $0.597176  	\times 10^{-9}$ \\ 
12   & 1 & $ 0.108387  	\times 10^{-4}$ & & 
36   & 0 & $0.464494  	\times 10^{-9}$ \\ 
13   & 1 & $ 0.582574  	\times 10^{-5}$ & & 
37   & 0 & $0.435966  	\times 10^{-9}$ \\ 
14   & 0 & $ 0.504688  	\times 10^{-5}$ & &
38   & 1 & $0.400227  	\times 10^{-9}$ \\
15   & 0 & $ 0.195936  	\times 10^{-5}$ & &
39   & 1 & $0.367349  	\times 10^{-9}$ \\ 
16   & 0 & $ 0.879431  	\times 10^{-6}$  & & 
40   & 0 & $0.258806  	\times 10^{-9}$ \\ 
17   & 1 & $ 0.696052  	\times 10^{-6} $ & & 
41   & 0 & $ 0.179591  	\times 10^{-9} $\\ 
18   & 0 & $ 0.423968  	\times 10^{-6} $ & &
42   & 1 & $ 0.164412  	\times 10^{-9} $\\ 
19   & 1 & $ 0.360942  	\times 10^{-6} $ & &
43   & 1 & $ 0.125568  	\times 10^{-9} $\\ 
20   & 0 & $ 0.110189  	\times 10^{-6} $ & &
44   & 1 & $ 0.113751  	\times 10^{-9} $\\ 
21   & 0 & $ 0.707359  	\times 10^{-7} $ & &
45   & 0 & $ 0.838048  	\times 10^{-10} $\\ 
22   & 1 & $ 0.604211  	\times 10^{-7} $ & &
46   & 0 & $ 0.475232  	\times 10^{-10} $\\ 
23   & 1 & $ 0.427386  	\times 10^{-7} $ & &
47   & 0 & $ 0.398470  	\times 10^{-10} $\\ 
24   & 0 & $ 0.372170  	\times 10^{-7} $ & &
48   & 0 & $ 0.355924  	\times 10^{-10} $ \\ 
& & & &
49   & 1 & $ 0.319369   \times 10^{-10}$\\ 
\hline
\end{tabular} 
\caption{{\protect{\footnotesize Diagonal elements of ${\cal A}$ for
 the square lattice 
for the $n = 48$ approximation}}}
\end{table}
}

\begin{table}
{\renewcommand{\arraystretch}{0.7} 
\begin{tabular}{|rclc|rcl|}
\hline
$i$ & $s_i$ & ~~$a_i$ & & $i~~$ & $s_i$ & ~~~~$a_i$ \\ \hline
~~~~~1    & 0 & 1.0 & &
~~~~~ 11   & 1 & $1.71216	\times 10^{-9}$ \\ 
2    & 1 & $0.68397$ & &
12   & 0 & $6.35801 \times 10^{-10}$ \\ 
3 & 0 & $1.35405   \times 10^{-2} $  & &
13 &1 & $ 6.12314	\times 10^{-10} $ \\ 
4 &1 & $1.27829   \times 10^{-3} $  & &
14 & 0 & $9.24125	\times 10^{-11} $ \\ 
5 &0 & $1.12917   \times 10^{-4} $ & &
15 &0 & $1.19689	\times 10^{-11} $ \\
6 &1 & $9.20508	\times 10^{-6} $ & &
16 &1 & $1.03587	\times 10^{-11} $ \\ 
7 &0 &$ 1.02989	\times 10^{-6} $ & &
17 &1 &$ 4.37873	\times 10^{-12} $ \\ 
8 &0 & $1.35813	\times 10^{-7} $ & &
18 &0 & $3.65874	\times 10^{-12} $ \\ 
9 & 1 & $7.94582	\times 10^{-8} $ & &
19 &0 & $8.98123	\times 10^{-13} $ \\ 
10 & 0 & $9.57785	\times 10^{-9} $ & &
20 &1 & $1.39953	\times 10^{-13} $ \\ 
& & & & 21 &  0 & $6.92231 \times 10^{-14}$ \\ \hline
\end{tabular} 
\caption{{\protect{\footnotesize Diagonal elements of ${\cal A}_1 {\cal A}_2$ for the honeycomb lattice for the $n = 20$
approximation}}}
}
\end{table}


\begin{thebibliography}{19}

\bibitem{VarApprox78}
R.J. Baxter, {\it Variational Approximations for Square Lattice Models in
Statistical Mechanics},
J. Stat. Phys. {\bf 19} (1978) 461 -- 478 

\bibitem{Baxter80}
R.J. Baxter, {\it Hard hexagons: exact solution}, J. Phys. A {\bf 13} (1980)
L61 -- L70

\bibitem{BaxterCTM81}
R.J. Baxter, {\it Corner transfer matrices}, Physica {\bf 106A} (1981) 18 -- 27 

\bibitem{BaxBook82}
R.J. Baxter, {\it Exactly Solved Models in Statistical Mechanics}, Academic, 
London (1982)

\bibitem{BaxterCitnClassic90} 
R.J. Baxter, {\it Numerics, conjectures and exact results}, Citation Classics,
in Current Contents (Phys. Chem. \& Earth Science) {\bf 30} (1990) 24

\bibitem{BaxterEnting79}
R.J. Baxter and I.G. Enting, {\it Series Expansions from Corner Transfer 
Matrices: the Square Lattice Ising Model}, J. Stat. Phys. {\bf 21}
(1979) 103 -- 123

\bibitem{BaxterEnting80}
R.J. Baxter, I.G. Enting and S.K. Tsang, {\it Hard Square Lattice Gas}, 
J. Stat. Phys. {\bf 22} (1980) 465 -- 489 

\bibitem{BaxterTsang80}
R.J. Baxter and S.K. Tsang, {\it Entropy of Hard Hexagons},
J. Phys. A (Math. Gen) {\bf 13} (1980) 1023 -- 1030 

\bibitem{Calkin98} 
N.J. Calkin and H.S. Wilf, {\it The  number of independent sets in a grid graph},
SIAM J. Discrete Math. {\bf 11} (1998) 54 -- 60

\bibitem{FinchWeb}
S. Finch, private communication: {\it Hard Square Entropy Constant},
MathSoft. Inc. Web site  http://www.mathsoft.com/asolve/constant/square/square.html

\bibitem{Finch98}
S. Finch, {\it Several Constants Arising in Statistical Mechanics},
http://xxx.lanl.gov/ abs/math.CO/9810155

\bibitem{Engel90}
K. Engel, {\it On the Fibonacci number of an $m$ by $n$ lattice},
Fibonacci Quarterly {\bf 28} (1990) 72 -- 78

\bibitem{Enting77}
I.G. Enting, {\it Triplet order parameters in triangular and honeycomb Ising  
models}, J. Phys. A: Math. Gen. {\bf 10} (1977) 1737 -- 1743

\bibitem{Fisher59}
M.E. Fisher, {\it Transformations of Ising models}, Phys. Rev. {\bf 113} (1959)
969 -- 981

\bibitem{Fisher63a}
M.E. Fisher, {\it Perpendicular susceptibility of the Ising model}, J. Math. Phys. {\bf 4} 
(1963) 124 -- 135

\bibitem{Fisher63b}
M.E. Fisher, {\it Lattice Statistics - A Review and an Exact Isotherm for a Plane 
Lattice Gas}, J. Math. Phys. {\bf 4} 
(1963) 278 -- 286

\bibitem{Gaunt67}
D.S. Gaunt, {\it Hard-Sphere Lattice Gases.II Plane-Triangular and Three-
Dimensional Lattices}, J. Chem. Phys. {\bf 46} (1967) 3237 -- 3259

\bibitem{GauntFisher65}
D.S. Gaunt and M.E. Fisher, {\it Hard-Sphere Lattice Gases. 
I. Plane-Square Lattice},
J. Chem. Phys. {\bf 43} (1965) 2840 -- 2863

\bibitem{Joyce88}
G.S. Joyce, {\it On the hard-hexagon model and the theory of modular functions},
  Phil. Trans. Roy. Soc. London {\bf A 325} (1988) 643 -- 702

\bibitem{KramersWannier41}
H.A. Kramers and G.H. Wannier, {\it Statistics of the Two-Dimensional Ferromaget. Part I},
Phys. Rev. {\bf 60} (1941) 252 -- 262 

\bibitem{Markley81}
N.G. Markley and M.E. Paul, {\it Maximal measures and entropy for 
$Z^n$ subshifts of finite type},
in Classical Mechanics and Dynamical Systems, Dekker Lecture Notes,
ed. R.L. Devaney and Z.H. Nitecki {\bf 70} (1981) 154 -- 155

\bibitem{McKay}
B.D. McKay,  private communication (1996)

\bibitem{MetcalfYang78}
B.D. Metcalf and C.P. Yang, {\it Degeneracy of anti-ferromagnetic Ising lattices 
at critical  magnetic field and zero temperature},
Phys. Rev. B {\bf 18}  (1978) 2304 -- 2307

\bibitem{Milosevic89}
S. Milosevic, B. Stosic and T. Stosic, {\it Towards finding exact residual entropies 
of the Ising ferromagnets}, Physica A {\bf 157} (1989) 899 -- 906

\bibitem{PearceSeaton88}
P.A. Pearce and K.A. Seaton, {\it A Classical Theory of Hard Squares},
J. Stat. Phys. {\bf 53} (1988) 1061 -- 1072 

\bibitem{RunnelsCoombs66}
L.K. Runnels and L.L. Coombs, {\it Exact Finite Method of Lattice Statistics. 
I. Square and Triangular Lattice Gases of Hard Molecules},
J. Chem. Phys. {\bf 45} (1966) 2482 -- 2492

\bibitem{RunnelsCoombs67}
L.K. Runnels, L.L. Coombs and J.P. Salvant, {\it Exact finite method of Lattice Statistics.
II. Honeycomb Lattice Gas of Hard Molecules},
J. Chem. Phys. {\bf 47} (1967) 4015 -- 4020

\bibitem{Sykes65}
M.F. Sykes, J.W. Essam and D.S.Gaunt, {\it Derivation of Low-Temperature Expansions for
the Ising model of a Ferromagnet and an Anti-Ferromagnet}, J. Math. Phys.
{\bf 6} (1965) 283 -- 298

\end{thebibliography}
\end{document}